\newcommand{\cpp}{C\raisebox{.2ex}{\small{++}}}

\documentclass{article}
\title{Universal Object Oriented Languages and Computer Algebra}
\author{Alexander Yu. Vlasov\\
Federal Center for Radiology, IRH \\
\small 197101, Mira Street 8, St.$\!$--Petersburg, Russia\\
\small E-mail: Alexander.Vlasov@pobox.spbu.ru, alex@protection.spb.su
}
\date{}

\begin{document}
\maketitle

\noindent{\hrulefill}

{\hfill \small \em Extended abstract}

\begin{abstract}
  The universal object oriented languages made programming more simple and
efficient. In the article is considered possibilities of using similar methods
in computer algebra. A clear and powerful universal language is useful if
particular problem was not implemented in standard software packages like
REDUCE, MATHEMATICA, {\it etc}. and if the using of internal programming
languages of the packages looks not very efficient.

Functional languages like LISP had some advantages and traditions for
algebraic and symbolic manipulations. Functional and object oriented
programming are not incompatible ones. An extension of the model of an object
for manipulation with pure functions and algebraic expressions is considered.


\end{abstract}

\section{Introduction}\label{intro}
 Despite of existence of many interactive algebra--numerical systems
(MatLAB or MathCAD\footnote{Either can use MAPLE library})
and systems of computer algebra (MACSYMA, REDUCE\cite{Calcul},
MATHEMATICA\cite{Wolfram}, MAPLE, {\em etc}.), some kind of algebraic problems
related with numerous difficult and formal manipulation could not be simply
rewritten for computer. The systems of computer algebra very useful if the
class of problems under consideration is already has standard implementation.
It is symbolic integration and differentiation, rational and integer number
arithmetic with arbitrary precision\cite{Akritas}, {\em etc.} \@.

For implementation of new classes of problems the most of systems of computer
algebra ({\sf CA}) have his own programming languages. Sometime it can be very
universal and powerful languages like LISP. For example, many new packages was
implemented in REDUCE/RLISP by users.

On the other hand the programming of the new methods in a system of {\sf CA}
can be even more difficult, than using universal modern languages like
Pascal\cite{Wirth76}, {\cpp}\cite{CppPL}, {\em etc.} with new powerful
techniques like {\em data types} or {\em object oriented
programming} ({\sf OOP}).

In the paper are considered possible extensions of the {\sf OOP} to make
possible of simple implementations to systems of {\sf CA}. The paper
is continue a theme presented earlier at \cite{AIHENP96}.

\section{Object Oriented Programming and Computer Algebra}\label{OOP}

 The development of big and difficult pieces of software is often related
with proliferation of errors and lost of clarity. The problem of effectiveness
of programming was one of central point for creation structural languages
like Pascal\cite{Wirth76} and new generation of object oriented languages
like Java\cite{Java}. The LISP, traditional for {\sf CA}, is very universal,
but the structure of programs is too complicated.

\medskip

The {\sf OOP} has possibility of understanding representation of structures
in {\sf CA} \cite{AIHENP96}. For example it is possible to define {\em Module}
\footnote{Here it is {\em additive Abelian group}, unit is called ``zero''}:
{\obeylines
Module = \underline{\bf Object};
\quad \underline{\bf operation} $+$ (A,B : Module) : Module;
\quad \underline{\bf operation} $-$ (A,B : Module) : Module;
\quad \underline{\bf operation} $-$ (A : Module) : Module;
\quad \underline{\bf const} Zero : Module;
\underline{\bf end}; \{ Module \}
}
\noindent Then definition of {\em Ring} can exploit {\em inheritance} in
{\sf OOP}:
{\obeylines
Ring = \underline{\bf Object}(Module)
\quad \underline{\bf operation} $*$ (A,B : Ring) : Ring;
\quad \underline{\bf operation} $/$ (A,B : Ring) : Ring;
\quad \underline{\bf function} Inversion(A : Ring) : Ring;
\quad \underline{\bf const} Unit : Ring;
\underline{\bf end}; \{ Ring \}
}
It is conception of an abstract basic type. They are algebra, module, group,
field, ring {\em etc.}. The abstract types do not contain a data. Other
objects contain data. It is integer, real or complex numbers, quaternions,
$n \times n$ matrices, {\em etc.}:
{\obeylines
 Quaternion = \underline{\bf Object}(Algebra)
  \quad Data : \underline{\bf array} [0..3] \underline{\bf of} Number;
  \quad \underline{\bf function} Norm : Number;
  \quad \underline{\bf operation} $+$ (A,B : Quaternion) : Quaternion;
  \quad \underline{\bf operation} $*$ (A,B : Quaternion) : Quaternion;
  \qquad $\cdots$
\underline{\bf end}; \{ Quaternion \}
}
\medskip

There is a difficulty because in {\sf CA} it is necessary to work with
analytical expressions associated with each of the object. We can have
possibility to write either $z := 2*i$, or $z := x + i\,y$. In the
work \cite{AIHENP96} already was discussed some kind of extensions to
the model of an object. It is some synthesis of functional and object
oriented programming.

\pagebreak
\noindent{\hrulefill}

{\hfill \small \em Comments to attached copy of slide presentation}

\section*{Slide 1}

 {\bf A scientific problem} could be resolved with using variety of different
ways. Here is represented using: {\em Standard systems of computer algebra},
{\em Universal programming languages} and traditional work with {\em pen and
piece of paper}.

$$\downarrow$$

 An application of universal programming languages like C++, Pascal, etc. is still
widely used not only for numerical calculation but for any scientific data
manipulation, including computer algebra. Here is represented an approach
with using {\em Object oriented programming} for computer algebra.

$$\downarrow$$

 An object oriented program for computer algebra consists of {\em Specific
structures and algorithms of computer algebra} together with {\em Standard
structures of the object oriented language}.

$$\downarrow$$

 It is convenient to merge both types of structures in some {\em Extension
of object oriented language for computer algebra}.

$$\downarrow$$

 Practical realization of such idea could be either {\em standalone translator}
for the language or {\em Translator CA $\rightarrow$ OOP}, i.e. convertor
of {\em a program for computer algebra} to {\em a program on one of widespread
object oriented languages (C++, JAVA, Delphi, etc.)}. The {\em code generator}
produces program on the OO language, standard algorithms of computer algebra
also can be included in {\em Libraries} in a standard format accessible for
the language. Next, the produced C++ program can be translated to executable 
module; if it is JAVA program it can be converted to bytecode for using in WWW 
applets; etc.\@.  It is the possibility to make computer algebra more effective, 
universal, fast, ``light'' and to include it as parts in other software projects.

\section*{Slide 2}

  Here is represented good agreement of principles of object oriented
programming with structures of standard algebraic models in mathematics,
{\it i.e.} inheritance of methods and properties between such {\em abstract}
objects as {\em Semigroup, Group, Module, Ring, Division ring, Field,
Algebra} and such descendant {\em ``actual''} objects as {\em Real and Complex
numbers or Polynomials}.

\section*{Slide 3}

  Here is emphasized one of specific property of mathematical language that
requires some extension of model of an object in object oriented programming.
  As a simple example is considered meaning of same variable $x$ in different
contexts:

$$\hbox{\em What does `x' mean?}$$

First meaning is traditional for standard programming language there `x' means
{\bf value} of variable $x$ and nothing else. But another meanings also are
essential and used in mathematics and computer algebra. The second meaning 
is `x' as some {\bf variable} $x$ of given type with value unknown or not
essential. The last, most difficult case is `x' as value of some {\bf function}
with (maybe partially) unknown arguments or expression like $x = a * y + 2$.

\noindent{\leaders\hbox{- }\hfill}

All three meaning are also used in {\em Pure functional programming} languages
there the functions with partially defined arguments are used and could be
described formally as some tree represented on the slide.

\noindent{\leaders\hbox{- }\hfill}

The trees are simply described by using usual {\em Structures in object
oriented languages} produced by {\em formal inversion of all arrows} in
diagram above.

\noindent{\leaders\hbox{- }\hfill}

But principles of type checking of an object oriented language should be
extended for such a case to make the new structure (an object with pointers
to argument and to method of evaluation of the function) compatible with
initial type of variable $x$. It require an {\em Extension of object 
programming: compatibility of 3 subtypes} mentioned above.

\end{document}